\documentclass[prl,aps,twocolumn,showpacs]{revtex4-1}
\usepackage{amsmath,amssymb}
\usepackage{graphicx}
\usepackage{subfigure}
\def\pcomma{$^,$}
\def\pitep{$^1$}
\def\pgwu{$^2$}
\def\ppnpi{$^3$}
\def\pacu{$^4$}
\def\pmisis{$^5$}

\begin{document}

\title{High-precision measurements of $\pi p$ elastic differential cross 
sections in the second resonance region}

\author{
I.~G.~Alekseev\pitep,
V.~A.~Andreev\ppnpi,
I.~G.~Bordyuzhin\pitep\pcomma\pmisis,
W.~J.~Briscoe\pgwu,
Ye.~A.~Filimonov\ppnpi,
V.~V.~Golubev\ppnpi,
A.~B.~Gridnev\ppnpi,
D.~V.~Kalinkin\pitep,
L.~I.~Koroleva\pitep,
N.~G.~Kozlenko\ppnpi,
V.~S.~Kozlov\ppnpi,
A.~G.~Krivshich\ppnpi,
B.~V.~Morozov\pitep,
V.~M.~Nesterov\pitep,
D.~V.~Novinsky\ppnpi,
V.~V.~Ryltsov\pitep,
M.~Sadler\pacu,
B.M.~Shurygin\pitep,
I.~I.~Strakovsky\pgwu,
A.~D.~Sulimov\pitep,
V.~V.~Sumachev\ppnpi,
D.~N.~Svirida\pitep,
V.~I.~Tarakanov\ppnpi,
V.~Yu.~Trautman\ppnpi,
R.~L.~Workman\pgwu
\\
\vspace*{0.1in}
(EPECUR Collaboration and GW INS Data Analysis Center)
\vspace*{0.1in}
}
\affiliation{
\pitep Institute for Theoretical and Experimental Physics, Moscow, 117218, 
	Russia}
\affiliation{
\pgwu The George Washington University, Washington, DC 20052, USA}
\affiliation{
\ppnpi Petersburg Nuclear Physics Institute, Gatchina, 188300, Russia}
\affiliation{
\pacu Abilene Christian University, Abilene, Texas, TX 79699-7963, USA}
\affiliation{
\pmisis National University of Science and Technology ``MISiS", Moscow, 
	119049, Russia}

\date{\today}

\begin{abstract}
Cross sections for $\pi^{\pm} p$ elastic scattering have been measured 
to high precision, for beam momenta between 800 and 1240~MeV/c, 
by the EPECUR Collaboration, using the ITEP proton synchrotron. 
The data precision allows comparisons of the existing partial-wave 
analyses (PWA) on a level not possible previously. These comparisons 
imply that updated PWA are required.
\end{abstract}


\maketitle

Measurements of $\pi p$ elastic differential cross sections by the 
EPECUR group, at the ITEP 10~GeV proton synchrotron, have produced 
data of unprecedented precision for beam momenta from 800 to 1240~MeV/c
(2638 $\pi^+p$ and 4277 $\pi^-p$ data points). 
This energy range, which covers center-of-mass energies from 1560 to 
1800~MeV, was motivated by the search for a narrow structure associated 
with the pentaquark anti-decuplet~\cite{theta}, 
expected near 1.7~GeV. 

The data precision required to search for such a narrow structure has
produced cross sections capable of identifying both narrow 
resonance-like signals and cusps expected to appear at the thresholds
of opening production channels, such as $K\Lambda$ and $K\Sigma$.
The precision also greatly exceeds that of previously available cross 
sections, which were used to generate the Karlsruhe-Helsinki~\cite{kh} 
(KH) and Carnegie-Mellon-Berkeley~\cite{cmb} (CMB) fits, from which much 
of non-strange baryon spectrum was determined. This allows a comparison 
of the classical KH and CMB analyses, and the more recent GW 
results~\cite{dac}, at a level not possible with the existing database. 

Below, we first describe the experimental design and analysis. We then 
outline cases where a clear distinction exists between the new data and 
some of these older analyses.

\begin{figure}[th]
\centerline{\includegraphics[width=0.4\textwidth]{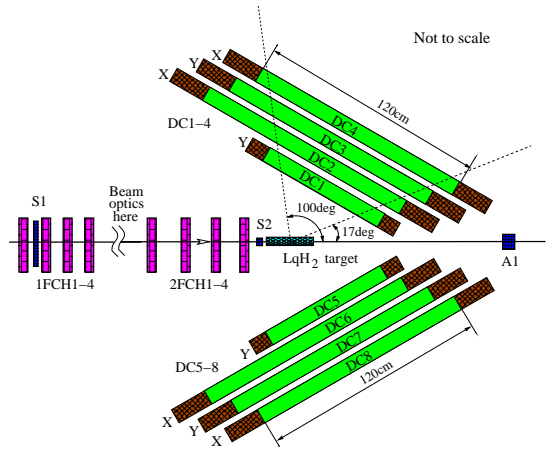}}
\caption{(Color on-line) The experimental setup (top view). {\bf 1FCH1-4} and 
	{\bf 2FCH1-4} --- 1-mm pitch proportional chambers, {\bf DC1-DC8} ---
	drift chambers, {\bf LH$_2$} --- liquid hydrogen target, {\bf S1}, 
	{\bf S2} and {\bf A1} --- trigger scintillation counters. 
\label{fig:setup}}
\end{figure}

The layout of the experiment~\cite{epecur} is shown in Fig.~\ref{fig:setup}. 
This is a two-arm non-magnetic spectrometer placed in the second
focus of a universal high resolution secondary beam line of the ITEP proton 
synchrotron. The first focus of the beam line is equipped by a set
of four 2-coordinate proportional chambers ({\bf 1FCH1-4} in Fig.~\ref{fig:setup})
with 1 mm pitch, which allow the tagging of each beam particle with its 
momentum with the precision about 0.1\%. Similar set of proportional chambers 
({\bf 2FCH1-4}) is placed in the second focus in front of the target. Beam 
size ($\sigma$) at the target is 5.5~mm and 3.5~mm in the horizontal and 
vertical planes correspondingly. ``Magic" (argon-isobutane-freon) gas 
mixture is used in proportional chambers. Beam tests showed better than 
99\% efficiency. The liquid hydrogen reservoir of the target is made of 
mylar and has 40~mm in diameter and about 250~mm in length along the beam. 
The reservoir is placed in a vacuum-tight 80~mm diameter beryllium outer 
shell with a mylar covered window on the beam entrance flange. Scattered 
particles are measured by two symmetrical arms of drift chambers ({\bf DC1-8}) 
with hexagonal structure. Each arm consists of 4 chambers. Wires in 
odd-numbered chambers are horizontal, with even-numbered wires being vertical. 
Each chamber has 2 sensitive wire planes with a 17~mm pitch. The planes are 
shifted by half of the pitch. The two chambers closest to the target have a 
sensitive area $600 \times 400$~mm$^2$. Six other chambers have sensitive 
area $1200 \times 800$~mm$^2$. A gas mixture of 70\% Ar and 30\% CO$_2$ is 
used in the drift chambers. Beam tests showed better than 99\% single drift 
plane efficiency with 0.2~mm resolution for perpendicular tracks. 

Central beam momentum was calibrated with 0.1\% precision at three values:
1057, 1095, and 1297~MeV/c using protons of the internal accelerator beam 
elastically scattered on the beryllium target. The field of the last dipole 
magnet of the beam line is controlled by NMR, providing stability of the 
energy calibration. In addition to the pions, the beam contains also electrons 
(positrons), muons and protons (for the positive beam).  Contamination from 
other particles (kaons and anti-protons) is negligible.  Protons were rejected 
at the trigger level by time of flight between scintillator counters in the 
first and the second focuses. The residual proton contamination was checked 
using the difference between $pp$ and $\pi p$ elastic kinematics and was 
found to be less than 0.2\%. The contribution of electrons and muons was 
measured using gas Cherenkov counter and simulated using Geant4~\cite{Geant4}. 
The fraction of electrons (positrons) is about 3\% at 840 MeV/c falling approximately linear to 1.5\%
at 1240 MeV/c and the fraction of muons falls in this range from 6\% to 4\%.

A unique distributed DAQ system, based on the commercial 480~Mbit/s USB~2.0 
interface, was designed for the experiment~\cite{epecurdaq}. It consists of 
100-channel boards for proportional chambers and 24-channel boards for drift 
chambers, placed on the chamber frames. Trigger logic is capable of processing 
several trigger conditions activating different sets of detectors. DAQ features 
nearly dead-time-less operation and can process up to $10^5$~events per spill.
A soft trigger condition was used to acquire physics events:

\begin{equation}
	T = S_1 \cdot S_2 \cdot M_{1FCH} \cdot M_{2FCH} \cdot \overline{A_1} ,
\end{equation}

where $S_1$, $S_2$, and $A_1$ are signals from corresponding scintillation 
counters and $M_{1FCH}$ and $M_{2FCH}$ are fast signals from the proportional 
chamber blocks in the 1$^\mathrm{st}$ and the 2$^\mathrm{nd}$ focuses. Other 
trigger conditions with large prescale were used for beam position and 
luminosity monitoring. During data taking the momentum range was scanned with 
15~MeV/c steps in the central momentum of the beam, which is about one half 
of the momentum spread in each step.

\begin{figure}[th]
    \centering
    \includegraphics[width=0.5\textwidth,height=0.3\textwidth]{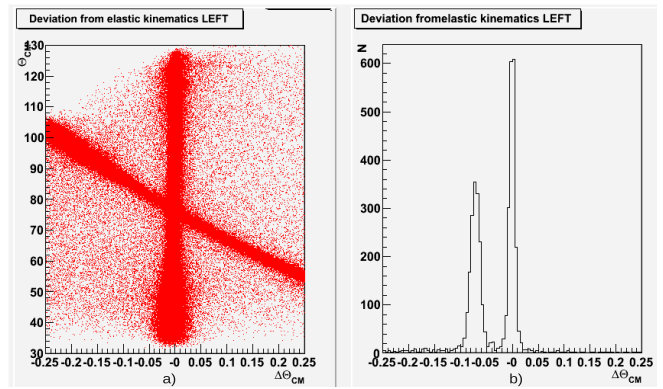}
    \caption{(Color on-line) 2-dimensional distribution over 
	the difference between calculated center of mass scattering angles for
	pion and proton, assuming that the pion goes to the left arm,
	$\Delta\theta_{CM}$ (absciss) and the scattering angle $\theta_{CM}$ (ordinate) 
	- (a) and its slice at $\theta_{CM} = 84^\circ$ (b).}
    \label{fig:QCM}
\end{figure}

Selection of the elastic events in this experiment is based on the angular 
correlation of pion and proton tracks. A single track is required in the beam 
chambers and both scattering arms. All of these tracks are required to form a 
common vertex inside the target and lie in a plane. A central of mass 
scattering angle $\theta_{CM}$ is calculated for both scattered particles 
under the assumption that the pion has scattered to the left. A distribution 
of the events over the difference between reconstructed scattering angles 
$\Delta\theta_{CM}$ and the scattering angle $\theta_{CM}$ is shown in 
Fig.~\ref{fig:QCM}a for one beam momentum setting. Two clusters are clearly 
seen. One corresponds to the pion scattered to the left (the assumption was 
correct) and the other corresponds to the pion scattered to the right (the 
assumption was wrong). A slice of the distribution for a one degree 
$\theta_{CM}$ interval $\theta_{CM} = 84^\circ$ is shown in 
Fig.~\ref{fig:QCM}b.  This figure also illustrates the amount of inelastic 
background, which was calculated and subtracted in each bin. Differential 
cross sections were calculated from the number of elastic events corrected 
for acceptance and chamber efficiency. Beam monitor is based on a special 
trigger, which ignores counter $A_1$, used as a veto in the main trigger. 
Numeric characteristics of the data sample are presented in 
Table~\ref{tab:stat}.

\begin{table}[th]
\caption{Parameters of the statistics presented}
\label{tab:stat}
\centering
\begin{tabular}{l|cc}
                                     & $\pi^-p\to\pi^-p$ & $\pi^+p\to\pi^+p$ \\
\hline
$\theta_{CM}$ angle range ($^\circ$) & $40 - 122$        & $40 - 122$  \\
Beam momentum range (GeV/c)          & $0.80 - 1.24$     & $0.92 - 1.24$     \\
Triggers accumulated                 & $1.25\cdot 10^9$  & $0.69\cdot 10^9$  \\
Elastic events                       & $2.24\cdot 10^7$  & $1.48\cdot 10^7$  \\
\end{tabular}
\end{table}

Main systematic error contributions are listed in Table~\ref{tab:sys}. We estimate 
total systematic uncertainty as sum in quadratures of the uncertanties in the table as 
2.6\%.

\begin{table*}[th]
\caption{Systematic errors}
\label{tab:sys}
\centering
\begin{tabular}{p{0.35\textwidth}|p{0.55\textwidth}|c}
Systematic error origin & Base for the estimation & Error \\
\hline
Beam pollution with electrons and muons & 
    Comparison between Monte-Carlo and Cherenkov counter measurements & 
        1\% \\
Luminocity normalization & 
    Comparison of elastic events yield for all angles in the overlapping 
    momentum ranges & 
        2\% \\
Tracking efficiency and setup geometry & 
    Comparison between cross sections obtained for events with pion 
    hitting the left arm to those with pion hitting the right arm & 
        1\% \\
Monte-Carlo simulations of the acceptance & 
    Comparison between two independent acceptance simulations used & 
        0.8\% \\
Various cuts used in the analysis &
    Dependence of the event yield from the cut &
        0.5\% \\
\end{tabular}
\end{table*}

In order to search for a narrow structure, data with a fine energy grid 
and high precision are required, and have been achieved. These cross 
sections have placed far higher constraints on existing partial-wave
analysis (PWA) than any previous experiment. As a result, angular 
structures are extremely well defined and clearly differentiate 
between the classic analyses of the KH~\cite{kh}, 
CMB~\cite{cmb}, and GW DAC~\cite{dac} groups. 
Their inclusion in future PWA will help to discriminate between 
competing mechanisms for sharp structures, such as the proposed 
antidecuplet, S-P wave resonance interference~\cite{shklyar}, or 
possible threshold cusp effects~\cite{julich}.

\begin{figure*}[th]
\includegraphics[height=0.8\textwidth, angle=90]{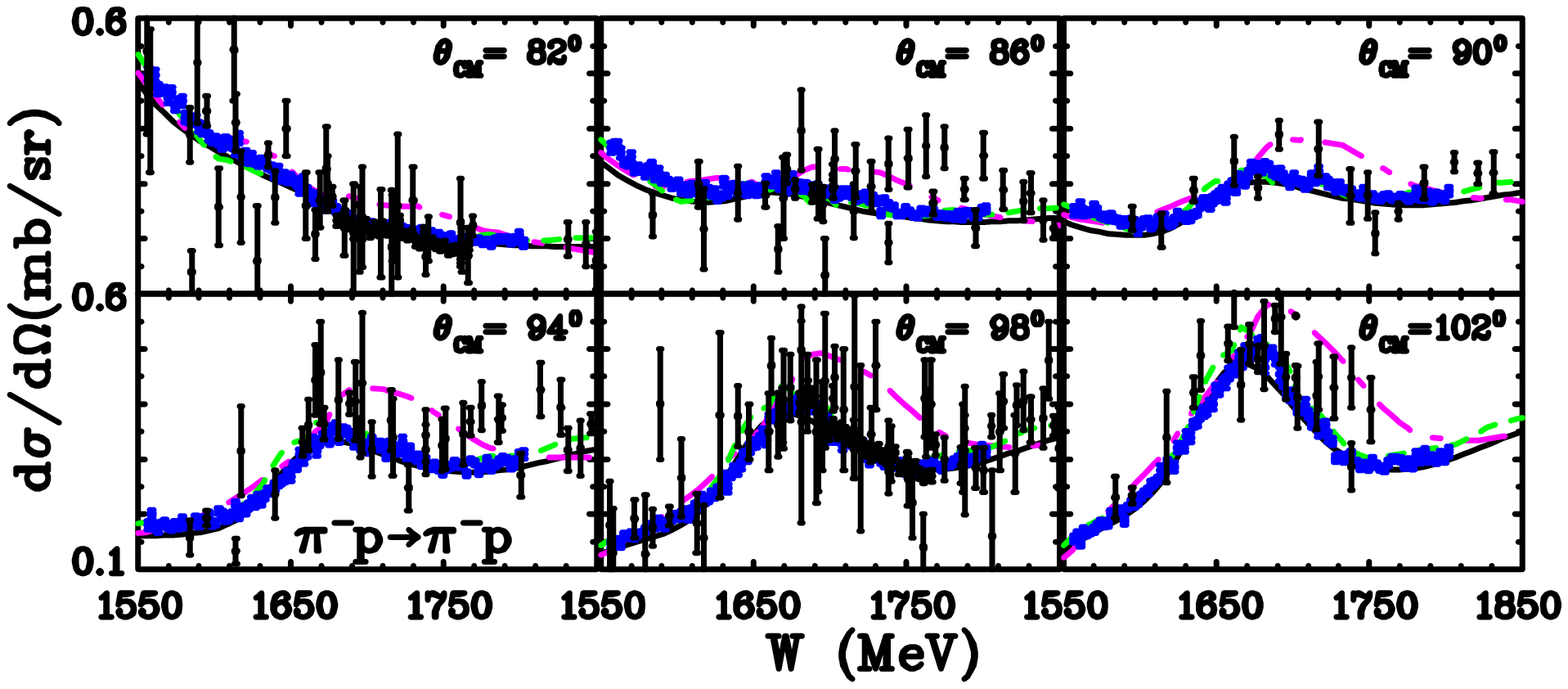}\hfill
\includegraphics[height=0.8\textwidth, angle=90]{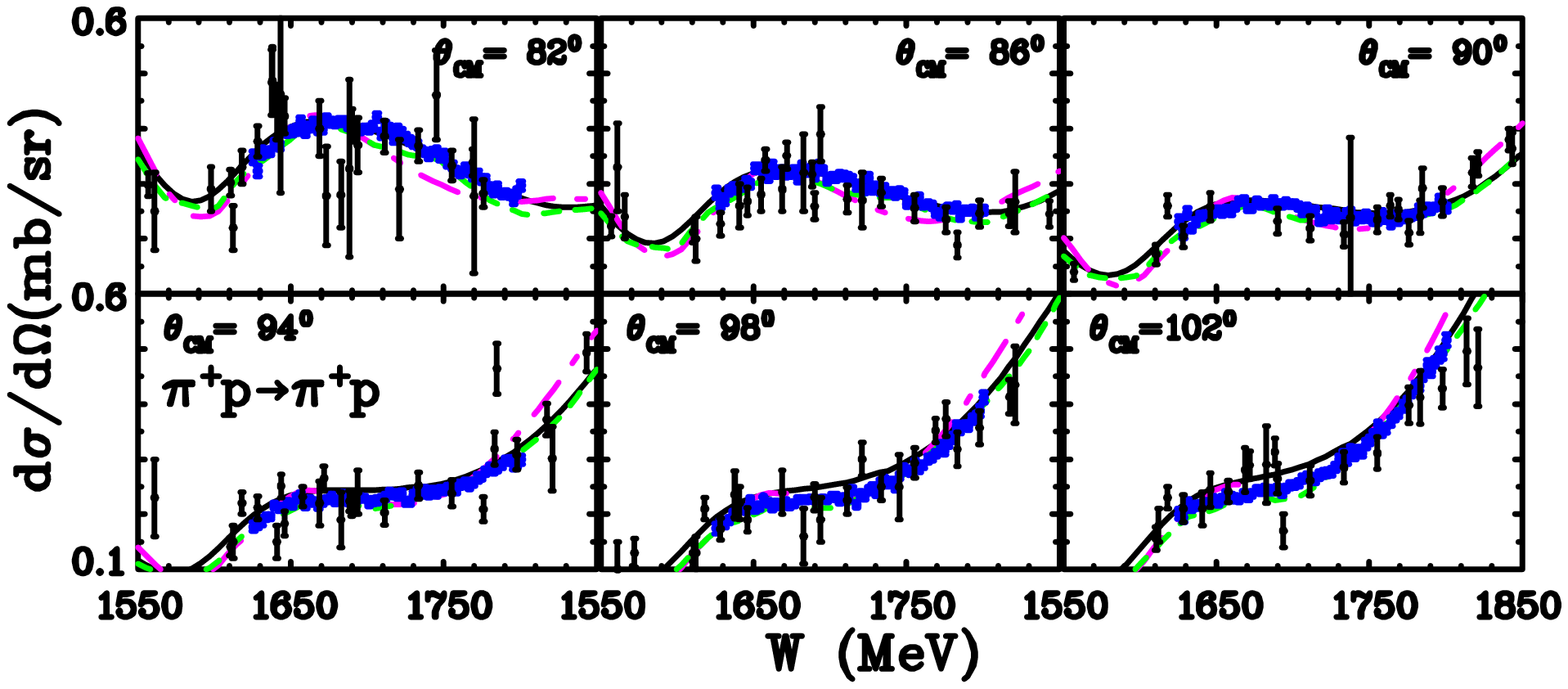}
\caption{(Color on-line) Excitation functions for selected angles around 
	90$^\circ$ in the center-of-mass frame, $\theta_{CM}$, for 
	$\pi^-p$ (top panel) and $\pi^+p$ (bottom panel) elastic 
	scattering. New EPECUR data (statistical errors only) are 
	plotted as blue filled circles 
	with previous measurements presented as black filled small 
	circles. (The data from earlier experiments are within bins of 
	$\Delta\theta_{CM}=\pm$1$^\circ$).  An existing GW INS DAC fit, 
	WI08~\protect\cite{dac}, is plotted with a solid black curve 
	while the older KA84~\protect\cite{KA84} and 
	CMB~\protect\cite{cmb} fits are plotted as magenta dash-dotted 
	and green dashed curves, respectively. \label{fig:gg03}}
\end{figure*}

In Fig.~\ref{fig:gg03}, we plot fixed-angle cross sections near 90 degrees
and compare with both the older datasets and the prediction based on fits 
to these older data. The higher precision now available can, in some cases, 
clearly select the older CMB and recent GW DAC fits over the KH fit (KA84). 
The earlier KH fit, KH80~\cite{kh}, fares somewhat better in these 
comparisons. Evidence for a possible sharp structure at more forward 
angles, near both a possible N(1685) resonance and the $K\Sigma$ 
threshold has been reported previously~\cite{nstar2013}. Such structures 
are not evident in any existing fit. The resonance hypothesis has also 
been tested in the fit of Ref.~\cite{gridnev}.

\begin{figure*}[th]
\includegraphics[height=0.3\textwidth, angle=90]{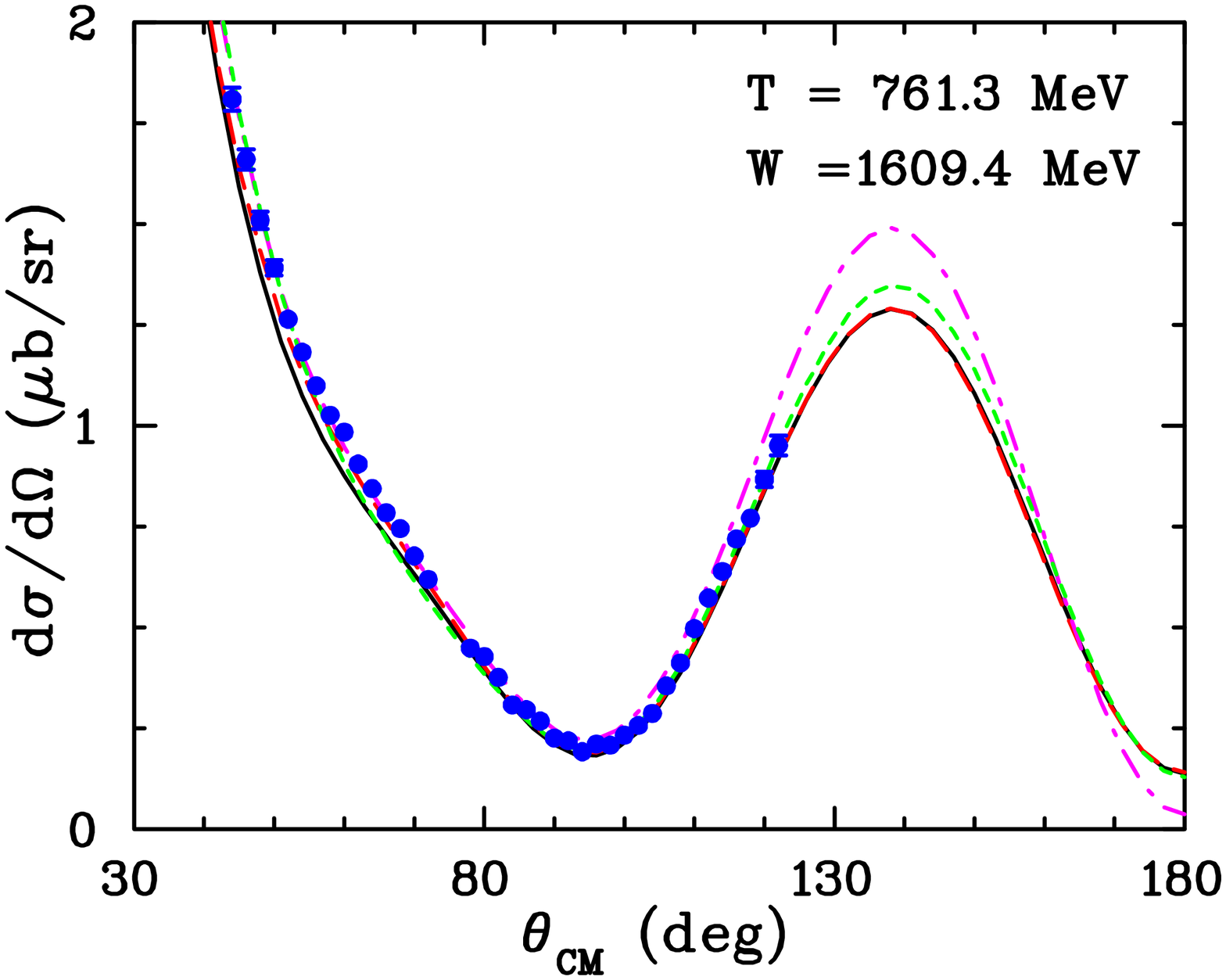}\hfill
\includegraphics[height=0.3\textwidth, angle=90]{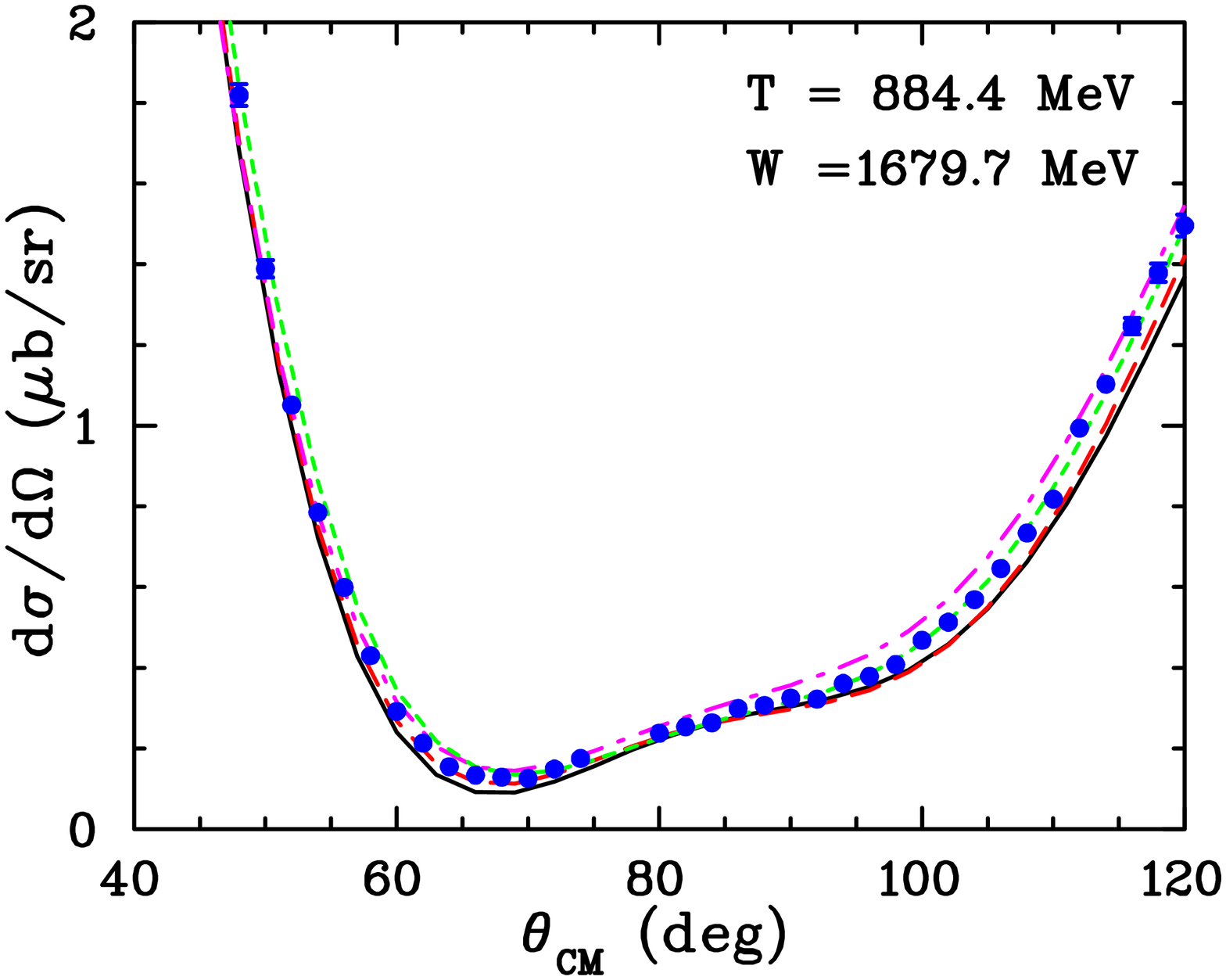}\hfill
\includegraphics[height=0.32\textwidth, angle=90]{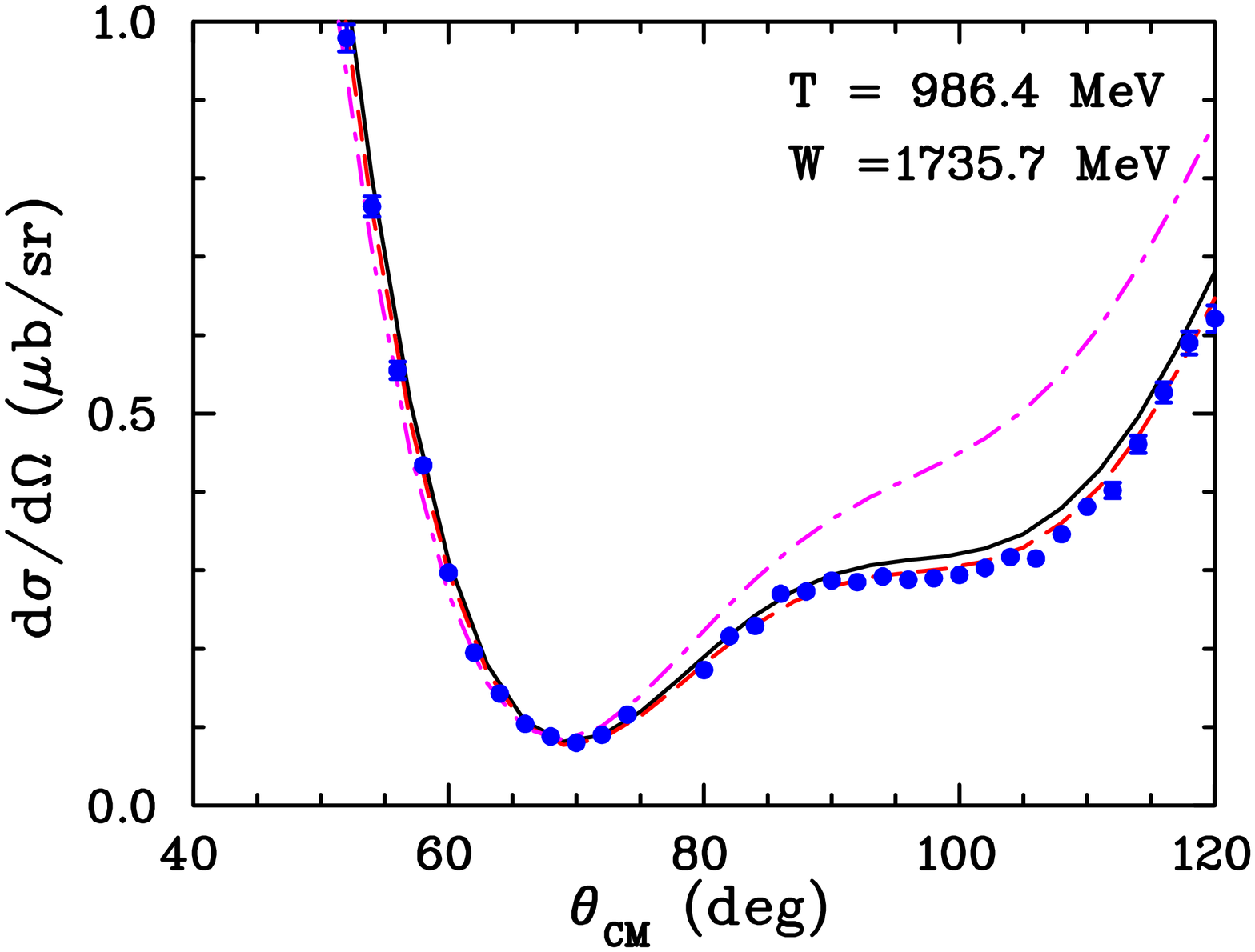}
\caption{(Color on-line) Differential cross sections for $\pi^-p$
	elastic scattering, $\theta_{CM}$, for selected energies.  
	Left: T=761.3~MeV, middle: 884.4~MeV, and right: 986.4~MeV. 
	New EPECUR data (statistical errors only) are plotted as blue 
	filled circles. For left 
	and middle panels, our recent prediction, WI08~\protect\cite{dac}, 
	is plotted with black solid curves while the older 
	KA84~\protect\cite{KA84} and CMB~\protect\cite{cmb} fits are 
	plotted as magenta dash-dotted and green dashed curves. For the 
	right panel, exploratory fit WI14 (with and without normalization 
	factor), including the new EPECUR data, is plotted with dashed 
	red and black solid curves respectively, while the older 
	KA84~\protect\cite{KA84} fit is plotted as a magenta dash-dotted 
	curve for comparison. \label{fig:gg04}}
\end{figure*}

In Fig.~\ref{fig:gg04}, we plot several angular distributions to give an 
alternate view of the fits/predictions versus data. From the left and 
central panel, we see how similar the existing predictions are, in general, 
with differences magnified in the view of Fig.~\ref{fig:gg03}. In the right 
panel, the KA84 fit is compared to a preliminary GW DAC fit to the new data. 
The GW fit employs a searchable renormalization factor, which gives a 
chi-squared penalty determined by the overall systematic error~\cite{dac}. 
In this case, as opposed to the other energies displayed in 
Fig.~\ref{fig:gg04}, the renormalization produces a good description of
the data, but is rather large compared to the systematic uncertainty, 
suggesting a more detailed analysis may be required. A more careful 
analysis could involve multi-channel fits with analytically built-in 
thresholds for opening channels.  Work in this direction is planned, based 
on the J\"ulich model of pion-induced reactions~\cite{julich}.


The authors wish to acknowledge the excellent support of the accelerator 
group and operators of ITEP. This work was partially supported by 
Russian Fund for Basic Research grants 09-02-00998a and 05-02-17005a
and by the U.S. Department of Energy, Office of Science, Office of Nuclear 
Physics, under Award Number DE.FG02.99ER41110.



\begin{thebibliography}{99}
\bibitem{theta}D.~I.~Diakonov, V.~Yu.~Petrov, and  M.~V.~Polyakov, 
	Z.\ F.\ Physik\ A\ \textbf{359}, 305 (1997);
	R.~A.~Arndt \textit{et al.}, Phys.\ Rev.\ C\ \textbf{69}, 
	035208 (2004). 
\bibitem{kh}G.~H\"ohler, {\it Pion Nucleon Scattering}, Part 2,
        Landolt-Bornstein, Vol.9b, 1983.
\bibitem{cmb}R.~E.~Cutkosky {\it et al.}, Phys.\ Rev.\ D\ \textbf{20},
        2839 (1979); R.~E.~Cutkosky in {\it Proceedings of the 4th
        Conference on Baryon Resonances}, ed. N.~Isgur, (Toronto, 1983).
\bibitem{dac} R.~L.~Workman, R.~A.~Arndt, W.~J.~Briscoe, M.~W.~Paris,
        I.I. Strakovsky, Phys.\ Rev.\ C\ \textbf{86}, 035202 (2012);
        R.~A.~Arndt, W.~J.~Briscoe, I.~I.~Strakovsky, and
        R.~L.~Workman, Phys.\ Rev.\ C\ \textbf{74}, 045205 (2006).
\bibitem{epecur} I.~G.~Alekseev \textit{et al.}, Instrum.\ Exp.\ Tech. 
	\textbf{57}, 535 (2014);
        I.~G.~Alekseev \textit{et al.}, arXiv:0509032 [hep-ex].
\bibitem{Geant4} S.~Agostinelli \textit{et al.} (Geant4 Collaboration),
	Nucl.\ Instrum.\ Meth.\ A\ \textbf{506} 250 (2003);
	http://geant4.cern.ch/
\bibitem{epecurdaq} I.~G.~Alekseev \textit{et al.}, Nucl.\ Instrum.\ 
	Meth.\ A\ \textbf{578} 289 (2007).
\bibitem{shklyar}V.~Shklyar, H.~Lenske, and U.~Mosel, Phys.\ Rev.\
        C\ \textbf{87}, 015201 (2013).
\bibitem{julich}M.~D\"oring and K.~Nakayama, Phys.\ Lett.\ B\ 
        \textbf{683}, 145 (2010).
\bibitem{KA84}R.~Koch, Z.\ Phys.\ C\ \textbf{29}, 597 (1985).
\bibitem{nstar2013} I.G.~Alekseev et al, Int.\ J.\ Mod.\ Phys.\ Conf.\ Ser.
    \textbf{26}, 1460076 (2014).
\bibitem{gridnev} A.~B.~Gridnev, Proceedings of the XVth International
	Conference on Hadron Spectroscopy (Hadron 2013), Nara, Japan, 
	Nov. 2013, Proceedings of Science (Hadron 2014) 099.
\end{thebibliography}
\end{document}